\documentclass[preprint,prl,aps,showpacs]{revtex4}
\usepackage{graphicx}
\usepackage{amsmath,amssymb}

\begin{document}

\title{Expansion of Solid Helium into Vacuum: the Geyser Effect}
\author{Robert E. Grisenti$^{1,2}$, Manuel K\"asz$^{2}$, J. Peter Toennies$^{1}$, Giorgio Benedek$^3$, and Franco Dalfovo$^4$}
\affiliation{$^1$Max-Planck-Institut f\"ur Str\"omungsforschung, Bunsenstr. 10, 37073 G\"ottingen, Germany} 
\affiliation{$^2$Institut f\"ur Kernphysik der J. W. Goethe Universit\"at, August-Euler-Str. 6, 60486 Frankfurt/Main, Germany}
\affiliation{$^3$Dipartimento di Scienza dei Materiali and INFM, Universit\`a di Milano-Bicocca, Via Cozzi 53, 20125 Milano, Italy} 
\affiliation{$^4$Dipartimento di Matematica e Fisica and INFM, Universit\`a Cattolica, Via Musei 41, 25121 Brescia, and CRS-BEC, Trento, Italy}
\begin{abstract}
The particle flux through a two micron diameter orifice into vacuum from a source chamber filled with solid $^4$He exhibits a striking periodic 
behavior similar to that of a geyser. 
This phenomenon is attributed to a periodic collapse of the solid inside the source induced by the accumulation of excess vacancies injected 
at the orifice. The flux-time curves agree well with a kinetic model and provide direct information on the diffusivity of vacancies in solid $^4$He.
\end{abstract}
\maketitle
%----------------------------------------------------------------------
Lattice vacancies are always present in an otherwise perfect crystal in thermal equilibrium because of the corresponding increase in entropy.
Regarded classically, these vacant sites are localized defects which move only occasionally in a thermally activated diffusion process.
In solid helium, however, as a consequence of
the light atomic mass and weak attractive interaction potential the particles execute zero-point oscillations which cover a large fraction 
of the cell size\cite{Andreev82}. 
Owing to the large amplitude of the zero-point motion tunneling of a vacancy into a neighbor lattice site is expected to occur at significant rates. Vacancies thus become delocalized entities which can be treated as elementary quasi-particle excitations of the crystal occupying a finite band of energy states\cite{Hetherington68,Andreev69,Guyer72}, which is a unique feature of quantum solids.

Despite several decades of intensive experimental effort the present understanding of the underlying mechanism of vacancy diffusion in quantum solids is still incomplete, as
demonstrated 
by the variety of conflicting data for the basic vacancy parameters (see \cite{Burns94,Simmons94} and references therein). Very likely this arises from the difficulties in the interpretation of results obtained by indirect methods, especially in the absence of a realistic model of a vacancy\cite{Galli03,Gov99}. Furthermore, the previous experiments afforded little control over the vacancy concentration, which drops exponentially with decreasing temperature and is thus probably too low to reveal significant quantum effects.

This letter reports on a quite surprising discovery related to the above issues. In the expansion of solid $^4$He into a vacuum system through a small pinhole orifice the particle flux exhibits a periodic sequence of Geyser-like bursts. The periods are in the range of seconds or minutes, depending on the pressure and temperature. The phenomenon is interpreted in terms of {\it excess} vacancies injected into the solid at the orifice which diffuse inwards eventually causing a periodic collapse 
of a macroscopic portion of the solid, thereby resetting the initial conditions. The observed temporal behavior is shown below to be closely related to the high mobility of the vacancies, and provides direct information on vacancy diffusion in solid $^4$He.

A standard molecular beam source in which the gas is replaced by solid $^4$He is used in these experiments. This simple arrangement represents an idealization of 
previous studies of plastic flow of solid $^4$He for the determination of vacancy diffusivity\cite{Zuev98,Berent98}. In the apparatus, shown in Fig. 1, the helium initially in the gas state is admitted into the source at a pressure $P_0$ and a temperature $T_0$ maintained by a liquid-He bath cryostat\cite{Grisenti03}. The temperature was determined from the He vapor pressure above the liquid helium bath. 
The pinhole orifice with cross section $S_{\rm ex}$ has a diameter $d = 2$ $\mu$m and a comparable length, and is located coaxially at the end of a long 1 mm-dia. copper tube with cross section $S_0$. After passing through a separately pumped source chamber the narrow beam reaches the detector chamber through a 2.5 cm-dia. aperture, located about 10 mm beyond the orifice. The beam flux is proportional to the detector chamber pumping speed times the measured pressure increase $\Delta P_{\rm D}$ in the detector chamber, which is between $\simeq 3.5\cdot 10^{-5}$ and  $\simeq 6\cdot 10^{-5}$ mbar.  

In experiments at a constant source temperature $T_0$ [Fig. 2(a)], the beam flux is measured as the source pressure $P_0$ is gradually increased stepwise from below to above the melting pressure $P_{\rm m}(T_0)$. At $P_0 < P_{\rm m}(T_0)$ the liquid flows at a constant rate steadily increasing with increasing $P_0$\cite{Grisenti03}. As soon as $P_0$ exceeds $P_{\rm m}(T_0)$ the flux exhibits bursts with period $\tau_0$, each burst being characterized by a sudden increase in intensity followed by a slowly decaying tail. The period $\tau_0$ is found to increase with increasing source pressure $P_0$ [see inset of Fig. 2(a)] and, as shown in Fig. 2(b) at a given temperature and pressure, is constant within $\lesssim 5\%$, even in experiments over several hours. 

The plots of the period $\tau_0$ as a function of pressure at constant temperature [inset in Fig. 2(a)] or of temperature at constant pressure (Fig. 3) show that the period vanishes at either the melting pressure $P_{\rm m}$ or the melting temperature $T_{\rm m}$. The critical points at which $\tau_0$ vanishes follow, within the experimental error, the melting curve (see inset of Fig. 3). 

Optical observations like those reported in Ref. \cite{Grisenti03} indicate clearly a narrow liquid jet at the nozzle exit for all of the source conditions considered here. 
This indicates that there must be a solid-liquid (s/l) interface with an effective cross section $S_{\rm sl}$ ($< S_0$) at some intermediate distance $r_{\rm sl}$ from the nozzle where the pressure approaches the melting pressure $P_{\rm m}$, as shown schematically in Fig. 4. In the liquid layer $r_{\rm sl} < r < 0$ the pressure obeys Bernoulli's equation\cite{Grisenti03}, with the average velocity of the liquid increasing from the s/l interface to the orifice exit inversely proportional to the reduction in cross section $S_{\rm ex}/S_{\rm sl}$. The exit liquid velocity is simply related to $\Delta P_{\rm D}$ by $u_{\rm ex} = C\Delta P_D V_{\rm at} /d^2 \sim 10^2$ m/s, where $V_{\rm at}$ is the atomic volume of the liquid and $C$ is an apparatus constant.

The near equality of the time averaged fluxes above [broken line in Fig. 2(a)] and below the melting point suggests a plastic flow of the solid inside the
source cylinder similar to that of a liquid\cite{Suzuki73-77}.  A crucial difference between the dynamics of the solid and liquid, however, is expected in the region where the flux cross section shrinks from the source cylinder area $S_0$ to 
$S_{\rm sl}$. In this region the plastic flow of the solid is ensured by two mechanisms involving the motion of defects: one is the plastic deformation allowed by the motion of dislocations; the other is the injection into the solid of vacancies generated at the s/l interface with a corresponding release of atoms from the solid into the liquid. Driven by their inward relaxation\cite{Chaudhuri99} the vacancies climb up the strong pressure gradient. Although many vacancies segregate at extended defects, e.g., at the axis of edge dislocations causing dislocation climbing\cite{Nabarro:book} a large fraction will be pinned in the upstream region at $r_0$ where the pressure gradient drops to zero and the pressure quickly approaches $P_0$ (see Fig. 4). While dislocation climbing will support a constant flow, the accumulation of excess vacancies around $r_0$ is held responsible for the periodic bursts in the beam flux intensity. A periodic behavior of the particle flux due to a stick-slip motion of the solid inside the source is ruled out since in this case one would expect a decrease of $\tau_0$ with increasing $P_0$\cite{Rozman96}, whereas in the present experiments the period increases with increasing source pressure [see inset of Fig. 2(a)].

To model the oscillatory part of the flow it is assumed that a percolative collapse, occurring at some critical vacancy concentration $X_{\rm v}^{\rm c}$, leads to a bleaching of vacancies and resets the initial conditions. By equating the total flux of the liquid at the orifice to that at the s/l interface, the time dependent liquid exit velocity can be written
as
\begin{equation}
u_{\rm ex}(t) = (S_{\rm sl}/S_{\rm ex}) [u_0 + V_{\rm at} j_{\rm v}(t)] \, ,
\label{eq1}
\end{equation}
where $u_0$ is the stationary velocity due to the plastic flow of the solid inside the source tube and the second term accounts for the time dependent vacancy current density $j_{\rm v}(t)$ at the s/l interface. $j_{\rm v}(t)$ consists of a diffusive and a drift term
\begin{equation}
j_{\rm v}(t) = -D_{\rm v} x_{\rm v}^{\prime}(r_{\rm sl},t) + u_{\rm v}x_{\rm v}(r_{\rm sl},t) \, ,
\label{eq2}
\end{equation}
where $x_{\rm v}(r,t)$ is the excess in the vacancy concentration with respect to the local equilibrium value in the thin solid layer of thickness $\ell = r_0-r_{\rm sl}$ (Fig. 4) and $x_{\rm v}^{\prime}(r,t)$ its space derivative. $x_{\rm v}(r,t)$ is determined by solving the kinetic equation in one dimension\cite{Wolfe_et_al:book}
\begin{equation}
\frac{\partial x_{\rm v}}{\partial t} = D_{\rm v} \frac{\partial^2 x_{\rm v}}{{\partial r}^2} - u_{\rm v} \frac{\partial x_{\rm v}}{\partial r} - \frac{x_{\rm v}}{\tau_{\rm r}} + G(r,t) \, .
\label{eq3}
\end{equation}
In Eqs. (2) and (3) the vacancy diffusion coefficient $D_{\rm v}$ is related to the mobility $\mu_{\rm v}$ through the Nernst-Einstein relation $D_{\rm v}=kT\mu_{\rm v}$, and $u_{\rm v}=\mu_{\rm v}F-u_0$ is the vacancy drift velocity with respect to the laboratory frame under the action of a force $F$. The inward relaxation of the lattice surrounding a vacancy is characterized by a local change $\Delta V_{\rm v} < 0$ in the vacancy volume and causes a net force field
$F\simeq -\Delta V_{\rm v} \nabla P$ on the vacancy driving it towards higher pressures. Thermal generation and recombination processes are included in Eq. (3) through the vacancy recombination time $\tau_{\rm r}$. The external generation function 
$G(r,t)= X_{\rm v}^0[-\delta(t)\theta(r-r_{\rm sl})\theta(r_0-r) + u_{\rm s}\delta(r-r_{\rm sl})\theta(t)]$, where $\theta$ is the Heaviside step function, consists of two terms. The first term models the sudden depletion of vacancies by an amount $X_{\rm v}^0$ occurring in the narrow region $r_{\rm sl}< r < r_0$ at time $t = 0$.
By assuming a depletion of all the vacancies, $X_{\rm v}^0$ is approximately given by an average equilibrium vacancy concentration of $X_{\rm v}^0 \sim 10^{20}$ cm$^{-3}$\cite{Zuev98}. The second term describes the subsequent vacancy injection at the s/l interface with a vacancy generation velocity $u_{\rm s}$\cite{Note}.

Equations (2) and (3) predict in good agreement with the experiment the temporal behavior of the flux within a single period. For simplicity, $D_{\rm v}$ in Eq. (2) has been assumed to be constant in the solid layer $r_{\rm sl}< r < r_0$. Figure 2(b) 
compares the calculated and experimental flux curves for $T_0 = 1.74$ K and $P_0 = 31$ bar. The best fit was obtained with $D_{\rm v} \simeq 1.3\cdot 10^{-5}$ cm$^2$/s (corresponding to $\mu_{\rm v} \simeq 5.4 \cdot 10^{10}$ s/g), $u_{\rm v} \simeq 2\cdot 10^{-3}$ cm/s, $\tau_{\rm r} \simeq 60$ s, and with $u_{\rm s} = 2 u_{\rm v}$\cite{Note}. The order of magnitude of $D_{\rm v}$, which is consistent with earlier measurements\cite{Zuev98}, is typical of a liquid, thereby indicating non-classical diffusion of vacancies in the solid. 
	
Since the liquid is incompressible the exit velocity $u_{\rm ex}$ is related to the pressure of the liquid at the s/l interface, $P_{\rm sl}$, by the Bernoulli equation 
$u_{\rm ex}\simeq (2 V_{\rm at} P_{\rm sl})^{1/2}/m$, where $m$ is the mass of a He atom. Thus, the oscillations in the detected signal $\Delta P_{\rm D}$ correspond directly to small oscillations of the pressure at the s/l interface $P_{\rm sl}$ [Fig. 2(a)], right-hand ordinate scale]. The oscillations of $P_{\rm sl}$ with respect to $P_{\rm m}$ are in turn a natural consequence of the variable concentration of vacancies in the solid which affects the local melting pressure. 

The excess critical vacancy concentration $x_{\rm v}^{\rm c} = X_{\rm v}^{\rm c}-X_{\rm v}^0$ causing the collapse of the solid can be estimated from the measured $\tau_0$. Within the present kinetic model the period $\tau_0$ is approximately given by $\tau_0 \approx \tau_{\rm v}(x_{\rm v}^{\rm c}/X_{\rm v}^0)(\ell /\ell_D)^{1/2}$, where 
$\tau_{\rm v}=4D_{\rm v}/u_{\rm v}^2$ and $\ell_D=(D_{\rm v} \tau_{\rm r})^{1/2}$ is the diffusion length. By assuming $\Delta V_{\rm v} \approx -60$ $\AA^3$\cite{Chaudhuri99}, 
it follows that $\ell \approx 40$ $\mu$m and hence $x_{\rm v}^{\rm c}/X_{\rm v}^0 \approx 10$. Since the integrated intensity over a full period is a measure of the amount of injected vacancies, the observed growth of the oscillation amplitude with increasing source pressure [Fig. 2(a)] suggests that at higher $P_0$ a larger excess vacancies concentration is needed to induce the collapse of the solid in the source. This is consistent with the expected decrease of the equilibrium vacancy concentration with increasing pressure\cite{Zuev98}.

In conclusion, the temporal behavior of the beam intensity in expansions of solid $^4$He into vacuum through a pinhole orifice provides a remarkably
direct way to explore vacancy diffusion and mobility. The orifice acts as a well defined source of vacancies, which gradually diffuse into the solid and accumulate in the region where the local pressure approaches the externally applied source pressure. Periodically, the concentration of vacancies in this region reaches some critical value at which the solid collapses, resetting the initial conditions. The extreme simplicity allows this type of experiment to be easily extended to a systematic study of vacancy 
diffusion in systems other than helium. A particularly intriguing case is represented by solid {\it p}-H$_2$, for which previous studies
suggest at $T\approx 10$ K the transition from a thermally activated to a purely quantum mechanical vacancy diffusion process\cite{Zhou89}.  
The present experiments are also ideally suited to address the exciting possibility of Bose-Einstein condensation of vacancies in bulk solid $^4$He\cite{Andreev69,Chester70,Leggett70}. Despite many attempts\cite{Meisel92}, only indirect evidence has so far been found for this ``supersolid" state\cite{Goodkind02}, perhaps because, if it exists, the equilibrium vacancy concentration is too small to make it detectable. In the present experiments a much larger vacancy concentration is achieved under non-equilibrium conditions, thereby enhancing the possibility of BEC in solid $^4$He. Indeed, the probable observation of a supersolid phase in solid helium confined in Vycor, a configuration that is likely to be more heavily populated with vacancies than bulk helium, has recently been reported\cite{Kim04} 

Useful discussions with L. Reatto, F. Pederiva, and W. Schr\"oter are gratefully acknowledged. We thank R. D\"orner for carefully reading the manuscript. R. E. G. and M. K. thank BMBF/GSI and DFG for financial support.

\newpage
%%%%%%%%%%%%%%%%%%%%%%%%%%%%%%%%%%%%%%%%%%%%%%%%%%%%%%%%%%%%%%%%%%%%%%%
%%  FIGURES
%%%%%%%%%%%%%%%%%%%%%%%%%%%%%%%%%%%%%%%%%%%%%%%%%%%%%%%%%%%%%%%%%%%%%%%
\begin{figure}
  \begin{center}
%     \leavevmode
     \includegraphics[width=10cm]{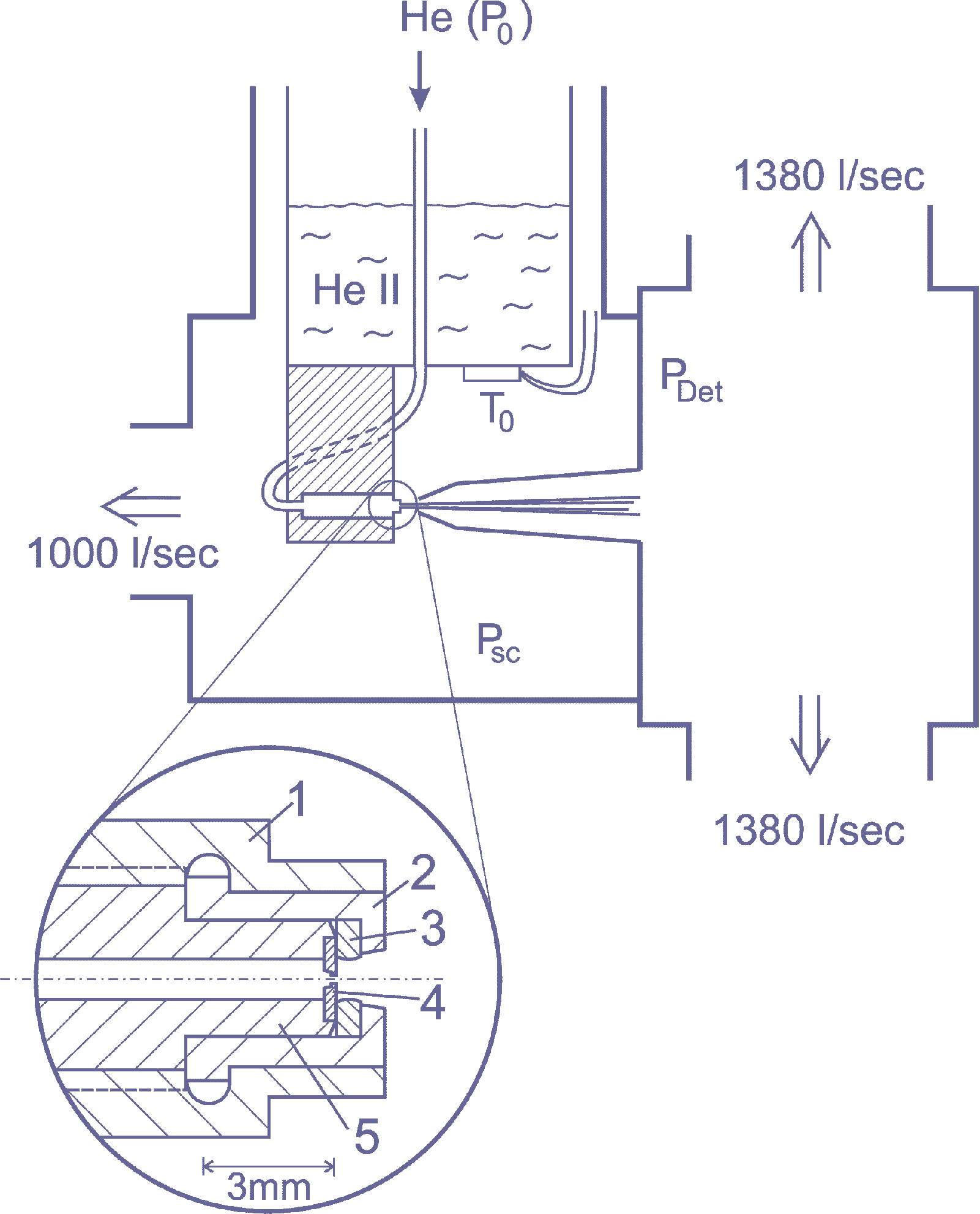} 
   \end{center}
   \caption{Schematic diagram of the source and vacuum system. Helium gas is admitted at a pressure $P_0$ into the source tube inside a copper block which is attached to a liquid-He cryostat at a temperature $T_0$. The liquid jet passes into the source chamber at a pressure of $\approx 10^{-6}$ mbar into the detector chamber. The inset shows a magnification of the source region. The 2 $\mu$m-dia. pinhole is centrally located in a thinned region of an 125 $\mu$m- thick 2 mm-dia. nickel platelet (4) at the end of a 1 mm inner dia. copper tube (5). The gold ring (3) serves as a pressure seal. The stainless steel retainer (2) is compressed by the outer stainless threaded cap (1).}
\end{figure}
\begin{figure}
   \begin{center}
%     \leavevmode
     \includegraphics[width=10cm]{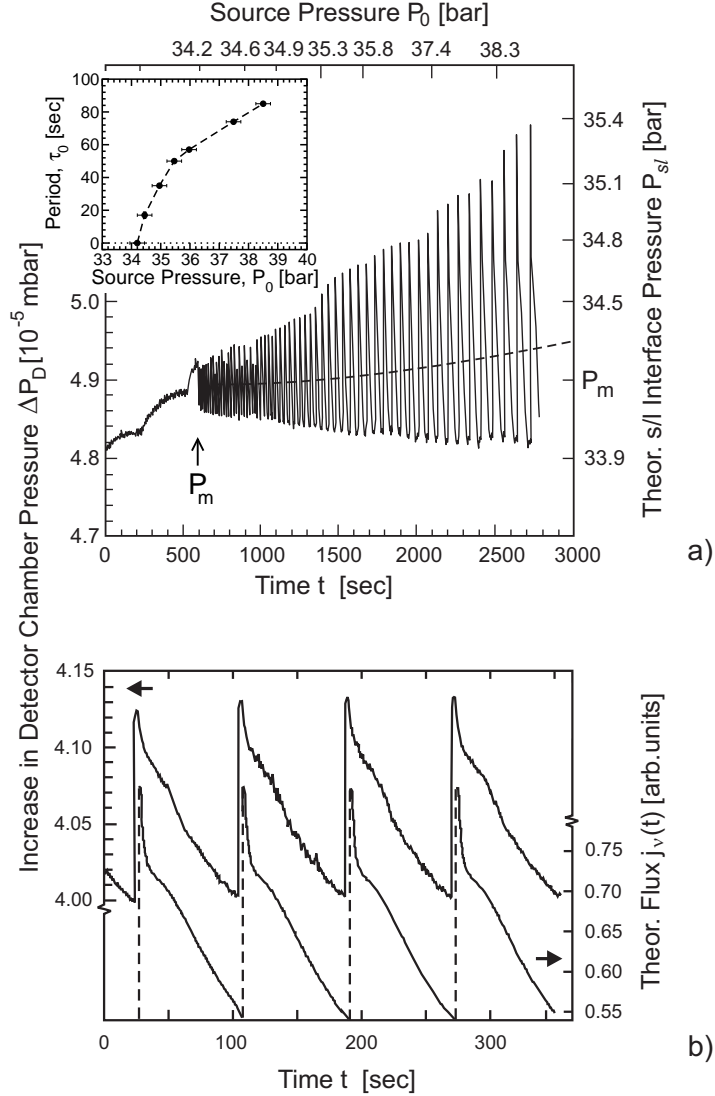} 
   \end{center}
   \caption{(a) The detector chamber pressure increase $\Delta P_{\rm D}$ as a function of time $t$ for $T_0 = 1.87$ K. The source pressure $P_0$ is gradually increased stepwise from below to above the melting pressure $P_{\rm m}(T_0)$ as indicated by the marks in the upper abscissa scale. The right-hand ordinate scale indicates the calculated pressure of the liquid at the solid/liquid interface corresponding to the observed beam intensity. The inset shows explicitly the oscillation period $\tau_0$ as a
function of $P_0$. (b) Four complete periods measured at $P_0 = 31$ bar and $T_0 = 1.74$ K are enlarged to show the recurring invariant shape with a constant period ($\tau_0 = 82$ s). The period and shape are very well reproduced by the pulse shapes, shown below the experiments, predicted by the excess vacancy transport theory Eq. (3). }
\end{figure}
\begin{figure}
   \begin{center}
%     \leavevmode
     \includegraphics[width=10cm]{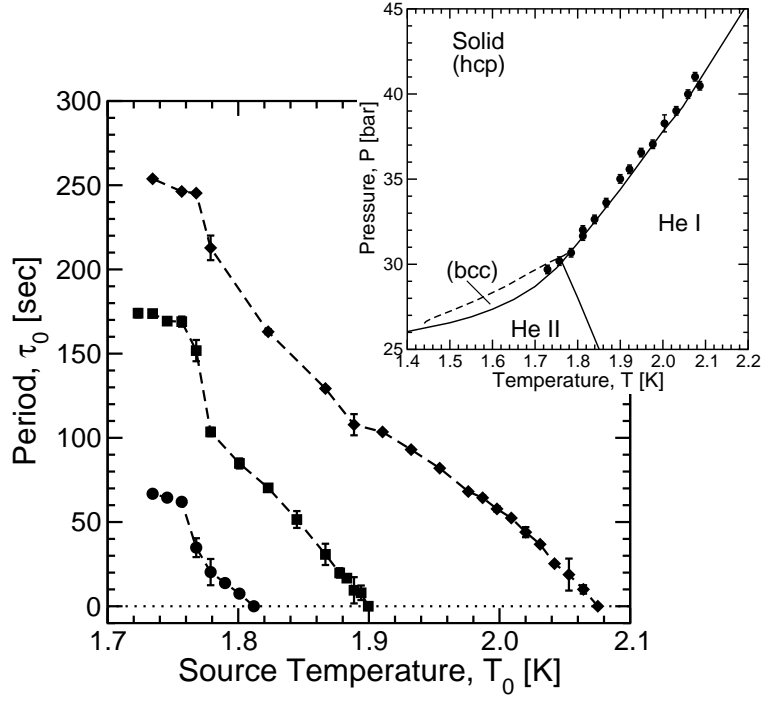} 
   \end{center}
   \caption{The measured oscillation period $\tau_0$ as a function of source temperature $T_0$ at $P_0=32$ bar (circle), 35 bar (square), and 41 bar (diamond). In the inset the points in the $P - T$ plane where the observed period $\tau_0$  vanishes all lie on the melting curve of the $^4$He phase diagram.}
\end{figure}
\begin{figure}
  \begin{center}
%    \leavevmode
    \includegraphics[width=10cm]{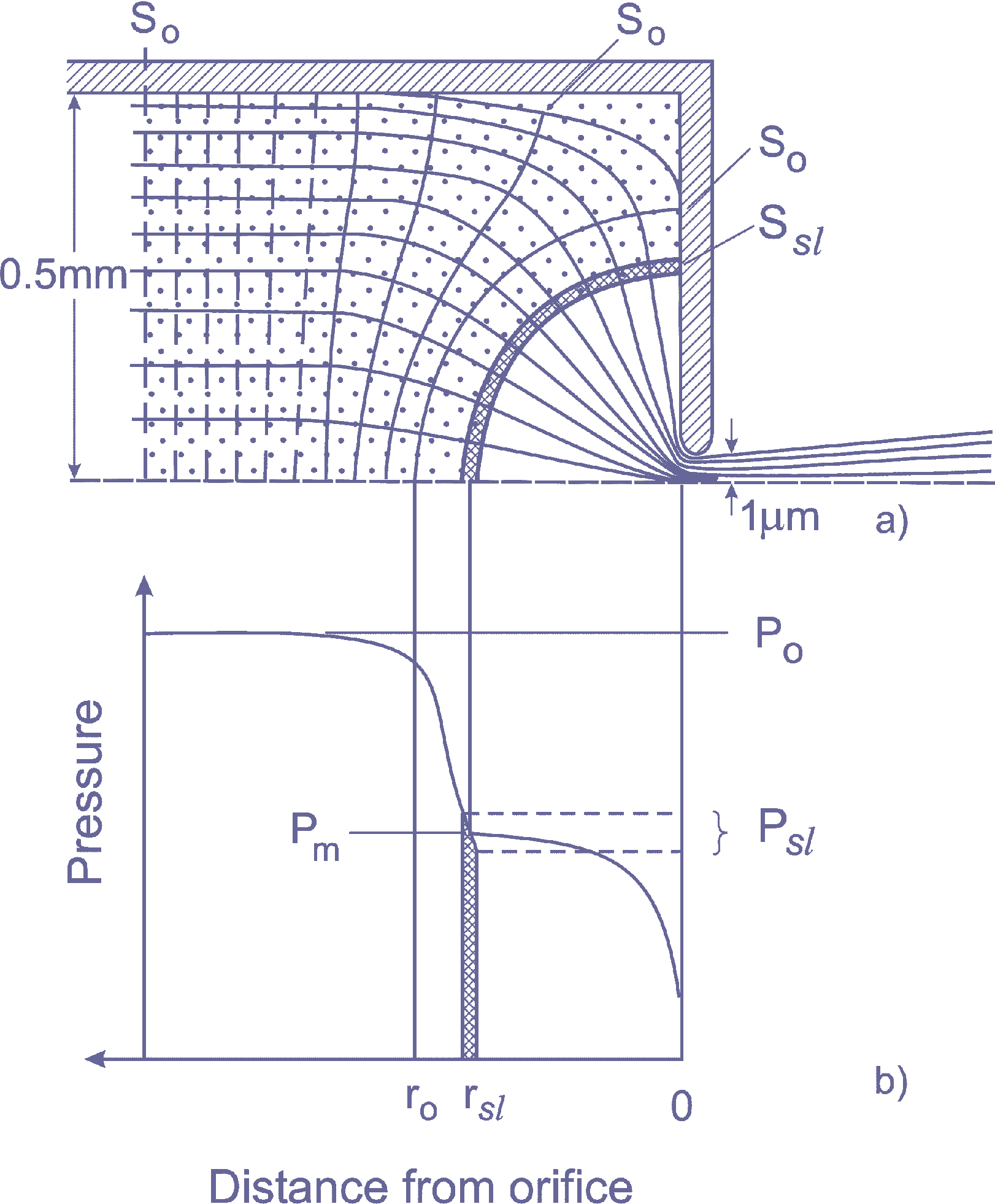}
  \end{center}
  \caption{(a) Schematic diagram showing the flow tubes near the orifice inlet and (b) the dependence of the pressure as a function of the distance $r$ from the orifice for the liquid ($r<r_{\rm sl}$) and for the solid in equilibrium with the liquid at the beginning and end of each pulse indicated by the cross-hatched region. In the solid layer $r_{\rm sl} < r < r_0$ the flux lines start converging towards the orifice producing a strong pressure gradient.}
\end{figure}
%%%%%%%%%%%%%%%%%%%%%%%%%%%%%%%%%%%%%%%%%%%%%%%%%%%%%%%%%%%%%%%%%%%%%%%
%%  END  FIGURES
%%%%%%%%%%%%%%%%%%%%%%%%%%%%%%%%%%%%%%%%%%%%%%%%%%%%%%%%%%%%%%%%%%%%%%%
%------------------------------------------------------------------------


\begin{thebibliography}{99}
\bibitem{Note}
The vacancy surface generation velocity $u_{\rm s}=\lambda \tau_{\rm s}^{-1}$ describes the injection of vacancies within a surface layer of thickness $\lambda$ as a result of the transfer of atoms from the solid to the liquid phase at a net rate $\tau_{\rm s}^{-1}$. In a stationary state $u_{\rm s}$ is balanced by the recombination and the sweeping away of vacancies by the inward pressure gradient. Fast surface recombination yields a vacancy depletion layer within $\lambda$. The integration over $0< r <\lambda$ in Eq. (3) for a stationary state, $x_{\rm v} \ll X_{\rm v}^0$, gives $u_{\rm s}\approx 2u_{\rm v} + \lambda/\tau_{\rm r}$, where the second (recombination) term can be neglected at high pressure gradients.
\end{thebibliography}
\end{document}